# Robust Wiener filter based time gating method for detection of shallow buried objects

Ali Gharamohammadi
Electrical Engineering Department
Sharif University of Technology
Tehran, Iran
aligharamohammadi@gmail.com

Fereidoon Behnia
Electrical Engineering Department
Sharif University of Technology
Tehran, Iran
behnia@sharif.edu

Arash Shokouhmand
Lane Department of Computer Science and Electrical Engineering
West Virginia University
Morgantown, USA
as0357@mix.wvu.edu

*Abstract*— In detecting shallow buried underground objects, reflected power from ground, i.e. ground surface clutter makes the task extremely difficult. In order to remove ground clutter, conventional methods in the literature are not as much effective as we need for objects buried detection in shallow depths. In this paper, a robust method, based on Time gating and wiener filtering, is proposed, which is very precise and effective in Ultra wideband (UWB) imaging. The problem with time gating method solely is that the timing window length for unknown target depths cannot be determined beforehand with sufficient accuracy. Imprecise window length selection removes parts of target signals along with the clutter and increases missed detection probability. This paper proposes an algorithm to circumvent this problem by first using a wiener filter for cancellation of ground clutter to a reasonable extent and pre detection of target positions by average similarity function (ASF). The time gating method is then used in the second step using the information provided from the first step for window length selection. The combination of the two steps provides better detection of shallow buried objects with less missed detection of targets.

## I. Introduction

Buried objects identification (BOI) problem is usually solved to detect targets such as pipelines, landmines, cables, treasure and mines. Detection of buried objects with a shallow depth has been a hot topic in recent researches. On the other hand, in applications such as landmine detection, detection of all buried objects is of crucial importance necessitating an efficient method, i.e. a method which detects all buried objects, even at the cost of some artifacts [1].

UWB signal is a novel method which is used in different fields of anomaly detection such as BOI [2-7], breast cancer imaging [8-9], detection in LPI radar [10] and gives better resolution and detection rate. Since UWB signal is narrow in time domain, better resolution can be obtained for positioning of buried objects compared to the previous methods. To detect shallow buried objects, ground surface reflection removal is important. In BOI problem, the reflected signal from ground surface is usually much stronger than that from buried objects which makes shallow buried objects identification (BOI) difficult [8], but the window length is not determined precisely. Conventional methods such as mean subtraction and Singular Value Decomposition (SVD) are not efficient for shallow buried objects problem [2]. Time gating method proposed in [2] is a new method in BOI.

There are several methods to reconstruct images from collected data. Back Projection (BP) and Reflected power method are conventional methods which are compared using Average Similarity Function (ASF) [11]. In this paper, ASF is implemented to reconstruct the images of buried objects and results of different methods of clutter cancellation are compared with using criteria. In proposed method to remove reflected signal from ground surface, the presented method can to determine time gating window exactly.

This paper presents a new algorithm for BOI based on time gating and wiener filter. The results of this algorithm shows better efficiency in comparison with time gating and wiener filter methods alone. In [12], wiener filter improved images as a noise reduction method. Here we use this method together with the time gating method to further improve the results achievable by each method alone.

The experimental data used in this paper is downloaded from the University of Georgia Tech website. The target placement scenario is rather complex covering so many buried objects.

The remaining parts of this paper are organized as follows. In section II, some of the conventional and new ground clutter removal methods are described. In section III, the simplified version of ASF to reconstruct images is described. In section IV, experimental setup is explained and previously proposed methods to remove ground clutter are compared using the experimental data. Section V concludes the paper

## II. Clutter cancellation

The ground surface reflection is a very important problem in BOI. Different methods have been traditionally used to remove ground reflection. In this paper, wiener filter is proposed for primary clutter cancellation in time gating method. An algorithm is implemented and compared with mean subtraction, SVD and Time gating.

### A. Mean subtraction

Mean subtraction is the simplest method to eliminate ground surface reflection from data. The following describes this method.

$$x'_{m,n} = x_{m,n} - \frac{1}{N*M} \sum_{j=1}^{N} \sum_{i=1}^{M} x_{i,j} \quad (1)$$

where $X_{m,n}$ is the received signal in the location (m,n), $X'_{m,n}$ is the modified version of $X_{m,n}$ and n, m are number of scan positions in x and y directions respectively [2].

In [13], this method is claimed to be the best one for eliminating ground surface reflection, But in this method target signal power also decreases and there is no difference between clutter and target signals [14]. In any position where ground surface clutter is different from the average, this method is not capable of removing clutter completely. In other words, ground surface clutter removal is not complete whenever its variance is high. In addition, in this method, the target space must be sparse compared to the clutter. In the problem of pipeline and tunnel tracking, this method attenuates target signal. Since in the mean signal, target is also present and its behavior is similar to that of clutter. In shallow buried object problem in which clutter must be eliminated completely for target detection, [2] indicates that this method is inefficient.

### B. SVD

SVD is a method used to remove ground surface reflection based on common singular vectors [15]. This method applies to received data in a line scan as described in the following

$$X_m = [x_{m,1}, x_{m,2}, ..., x_{m,N}] \quad (2)$$

where $X_m$ can be represented as follows [10]

$$X_m = \sum_{k=1}^{L} \sigma_k u_k v_k^H \quad (3)$$

Here, H denotes conjugate transpose, $\sigma_k$ is the kth singular value, $u_k$ is the kth left singular vector, $v_k$ is the kth right singular vector and L is the number of nonzero singular values. The first singular vectors of every sweep line project clutter signal and are discarded in this method. But, in practice, in the decomposed data, the first singular vectors are not the same and for this reason clutter and target signal are not separable completely [15]. After elimination of clutter singular vectors, $X_m$ is changed to the following

$$X'_m = \sum_{k=v+1}^{L} \sigma_k u_k v_k^H \quad (4)$$

### C. Time gating

In UWB signal, the ground surface reflection is discriminated from buried object reflection in an appropriate time gate. The length of the time domain signal is related to the signal bandwidth. Minimum depth of target can be calculated as follows

$$R = \frac{c}{2B\sqrt{\varepsilon_r}} \quad (5)$$

where c is light speed in free space, B is bandwidth of the signal and $\varepsilon_r$ is permittivity. R describes Minimum separable depth which is determined according to medium properties [11].

### D. Wiener Filter

Wiener filter is usually used to reduce the level of noise accompanying signals [12] considering that the additive noise is not correlated with the received signal from targets. In the UWB BOI problem for ground clutter reduction, the ground reflected signal is always present in the received data. The reflected signal from different targets are almost perpendicular in UWB imaging [11]. Considering different origins, the signals can be assumed independent. Due to the UWB nature, Independence plus zero mean implies orthogonally [16]. In this paper to remove ground clutter reflection, the reflected signals from target and ground clutter are also assumed to be perpendicular. Received signal is modeled as follows in Fig. 1. In this paper, the reference signal for wiener filter is ground reflected signal which is assumed to be the average of all obtained data.

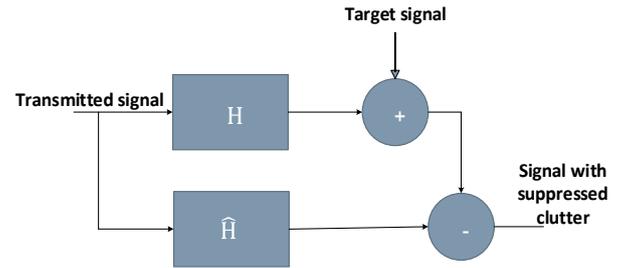

Fig. 1. Wiener filter block diagrm for clutter cancellation. In this method, H must be estimated to remove clutter from the received signal. The reflected signal from target and ground surface in this method are assumed to be orthgonal [11].

The ground clutter can be suppressed to a good extent in the reflected signal from the environment by estimating $\hat{H}$ and consequently discriminate between clutter and signal. The follwing experssion defines the operation of the system depicted in fig. 1.

$$T(n) = H.X_c + X_t - \hat{H}.X_c$$
$$E\{T^2(n)\} = E\{X_t^2(n)\} + E\{((H - \hat{H})X_c(n))^2\} \quad (6)$$

where $X_c$ and $X_t$ are the transmitted and the target signals respectively. T is the residual signal after clutter suppression. In this expression, minimization of the left side requires reduction of the 2nd part of the right side. Indeed, the distance between H and $\hat{H}$ is minimum when the second moment of T is minimized. This relation is showed as follow.

$$\min\ E\{T^2(n)\} \equiv \min\ H - \hat{H} \quad (7)$$

This point can be fulfilled by wiener filter. Wiener filter coefficients can be calculated by minimizing the second moment of T. For more details on wiener filter coefficients calculation see [16].

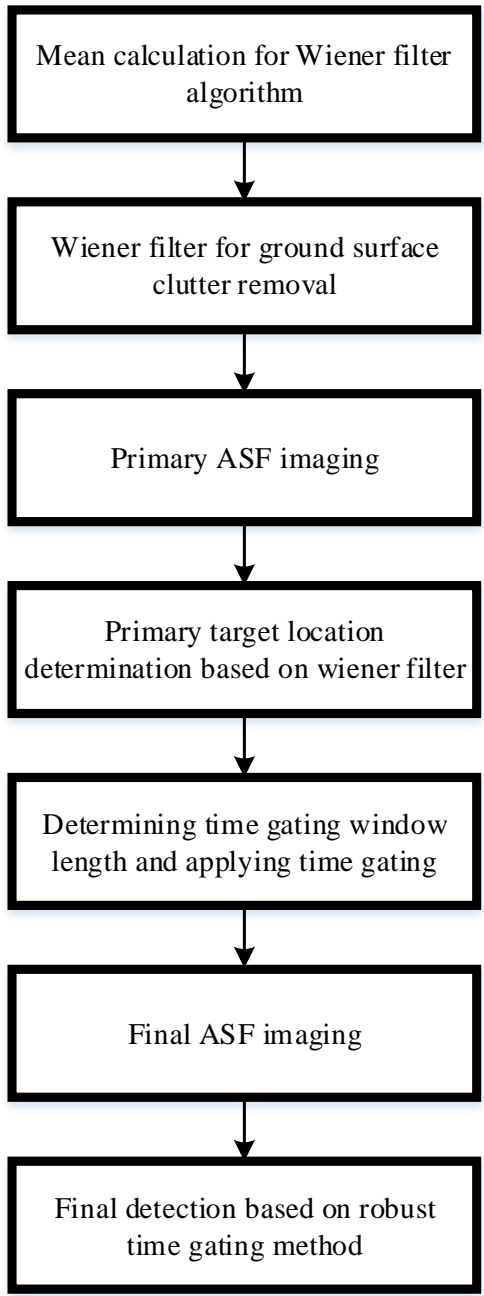

Fig. 2. Proposed algorithm for combined wiener filter and time gating for detecting buried objects. The primary detection is based on wiener filter and the final detection is based on time gating method. The result is a robust time gating algorithm for BOI.

*E. An algorithm based on Wiener filter and Time gating*

Time gating method is a precise method for removing ground surface clutter, but the window length in this method is not determined in prior. For this reason, time gating window can remove target signal in addition to the clutter. The proposed method in this paper is based on using wiener filter for primary estimation of environment and buried objects after ground surface clutter removal, the time gating window length is determined according to the position of the estimated targets in the environment. The proposed algorithm is depicted in Fig. 2.

The clutter removal window length is determined by the midpoint between the two sequential peaks in time domain signal. When the test scene is an empty ground with no targets, the second peak is the side lob of ground surface and this method removes ground clutter completely. In the case of buried objects it is the target signal. Fig. 3 shows the signal from two different positions where a target is present in one of them. The second peak shown in Fig. 3 is the target signal. In two different positions, the reflected signals from ground surface are not align because of the slight fluctuation of the ground surface.

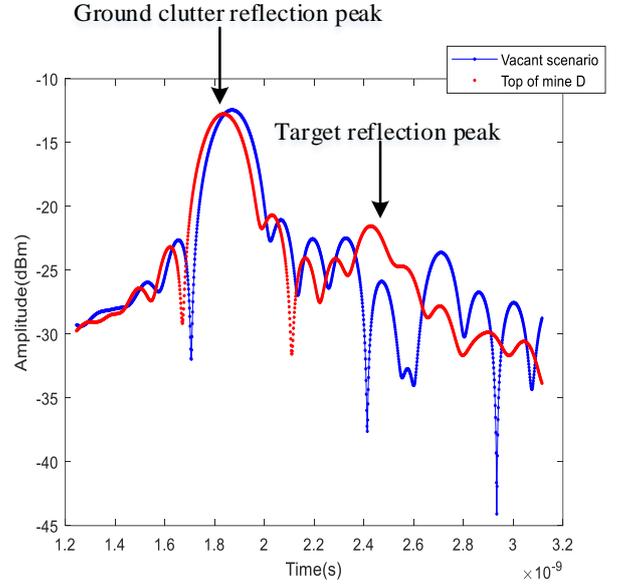

Fig. 3. Reflected signals form empty and non empty ground. The non empty case has a separated peak form ground clutter reflection peak in the time domain.

## III. IMAGING

In [8], the normalized correlation function is the criterion employed to calculate similarity of two signals. The function value for the same signals is 1 and for the completely perpendicular signals is 0. The received signals at the two different places, Q and K, can be defined as follows

$$Q = \begin{bmatrix} a_1 \\ a_2 \\ . \\ . \\ . \\ a_{M-1} \\ a_M \end{bmatrix}, K = \begin{bmatrix} b_1 \\ b_2 \\ . \\ . \\ . \\ b_{M-1} \\ b_M \end{bmatrix} \qquad (8)$$

where M is the length of each of the two vectors. The correlation function for the two discrete normalized signals is defined as follows

$$corr(Q,K) = \sum_{i=1}^{M} a_i b_i^* \qquad (9)$$

Where * sign represents the conjugate operation. The correlation function can be calculated in time or frequency domains. For this reason, the transform from frequency domain to time domain is not necessary [11].

The ASF is used to reconstruct images. ASF is based on correlation function and measuring similarity of a given position to others. This imaging method shows the position of objects from above. The primary function is as follows

$$E_{m,n} = \frac{1}{N^2} \sum_{i=1}^{N} \sum_{j=1}^{N} corr(x_{i,j}, x_{m,n}) \quad (10)$$

where $E_{m,n}$ denotes the ASF for the location (m,n) and $X_{m,n}$ is the received signal in the location (m,n). The above equation can be reduced to the following for easing calculations complexity

$$E_{m,n} = corr(avg(x), x_{m,n}) \quad (11)$$

## IV. EXPERIMENTAL RESULTS

### A. Data acquisition

Experimental data has been collected in a lab in Georgia Tech University. The experimental setup is described in the following. For more information see [1].

The system consists of an array of similar antennas. A vector network analyzer and a 3-D positioner are used to scan the region. The area size is 120*120 cm with fix height and 2 cm scan step in x and y axes. The transmit signal bandwidth is ca. 8 gigahertz with 20 megahertz frequency step. Permittivity of soil is 4. In this paper we only consider the case T1R1 which deals with the signal transmitted by the first transmitter and received by the first receiver. Different scenarios for data collection are implemented. The utilized scenario in this paper is depicted in Fig. 4.

In table 1, the relevant buried objects properties are mentioned. AT and AP are abbreviations for anti-tank and anti-personnel mines respectively.

### B. Simulation results of different methods

In the following the methods mentioned in the previous sections are compared according to their results in ASF images. In an ASF image, in any positions with no buried object, the similarity value is maximum, and in positions with buried objects the similarity value is minimum. As mentioned before, the area which is selected for image reconstruction is almost empty of buried objects. For example, in the scenario used in this paper, the area is almost 85% empty.

The results of mean subtraction and SVD methods are depicted in Fig. 5. For better dynamic range in figures have been set min/max of every figure. It is clear from these images that these method are not capable of suppressing ground clutter appropriately. As a results, many of the buried objects are not detected.

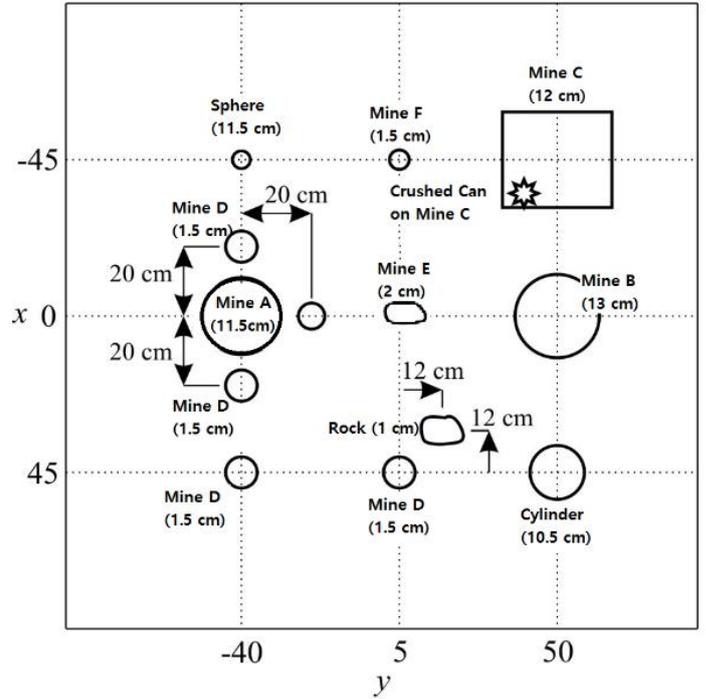

Fig. 4. Experimental setup. Name, location and depth of targets in x-y scene [1] are seen in the figure. The buried object materials are tabulated in table 1.

TABLE I. PROPERTIES OF BURIED OBJECTS IN EXPERIMENTAL SETUP[1].

| Target | properties | | |
|---|---|---|---|
| | *Type* | *Material* | *Dimensions(cm)* |
| Mine A | AT | Plastic | 22.2(D), 9.2(H) |
| Mine B | AT | Plastic | 24(D), 12(H) |
| Mine C | AT | Plastic | 31.2(L), 27.5(W), 11.3(H) |
| Mine D | AP | Plastic | 9(D), 4.5(H) |
| Mine E | AP | Plastic | 11.9(L), 6.4(W), 2(H) |
| Mine F | AP | Plastic | 5.6(D), 4(H) |
| Mine simulant | AP | Plastic | 7.5(D), 3.8(H) |
| Sphere | Buried clutter | Aluminum | 5.1(D) |
| Rock | Buried clutter | Rock | 12(L), 8(W), 7.5(H) |
| Crushed Can | Buried clutter | Aluminum | 8(D), 3(H) |
| Cylinder | Buried clutter | Nylon | 15.5(D), 7.6(H) |

The results of one step time gating and Wiener filtering methods are depicted in Fig. 6. The results are comparable in some aspects. The first is ability of finding almost all buried objects. All buried objects in the time gating method are identified except the rock in (33, 17) which is buried in 1 cm under the ground surface. The results also show that the Mine D in (45, -40), in image is almost suppressed and is difficult to detect. Indeed, time gating window length is not suitable and buried target signal is eliminated.

As can be seen, the wiener filter method also can detect nearly all buried objects expect the one in the right side of Mine A which is buried in (0, -20). Furthermore, wiener filter method has some artifacts as clear from the figure hence, time gating and wiener filter methods, although quite capable in detecting shallow buried objects compared to other methods mentioned in this paper, have some drawbacks. For this reason, in the following we combine the two methods by first running the Weiner filter method and running the time gating method for which, the window length is selected in accordance with the results obtained by application of wiener filter. The results of this algorithm is shown in Fig. 7. All buried objects are detected in this experimental scenario. However, some artifacts are still inevitable.

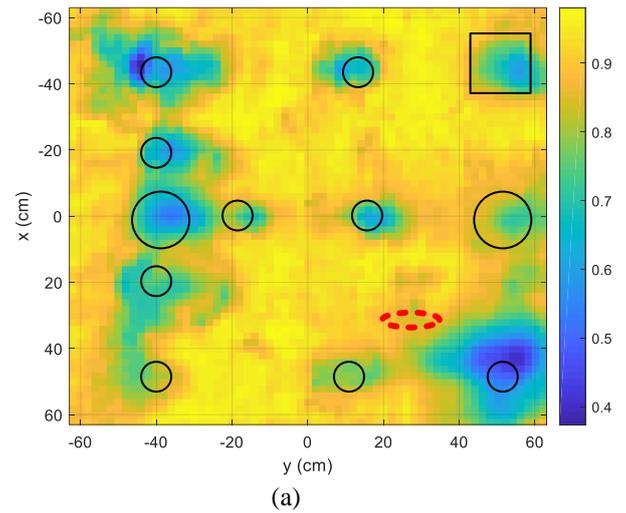

Fig. 5. Results of conventional methods. (a) mean subtraction (b)SVD. In these methods lot of buried objects are not detectable. Red dash lines show not detected objects.

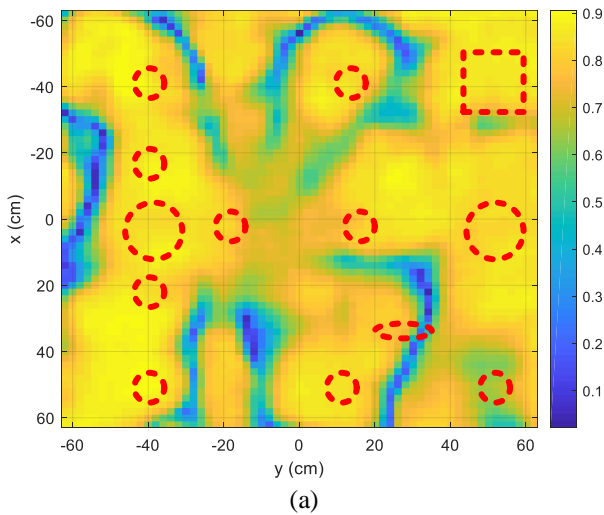

(a)

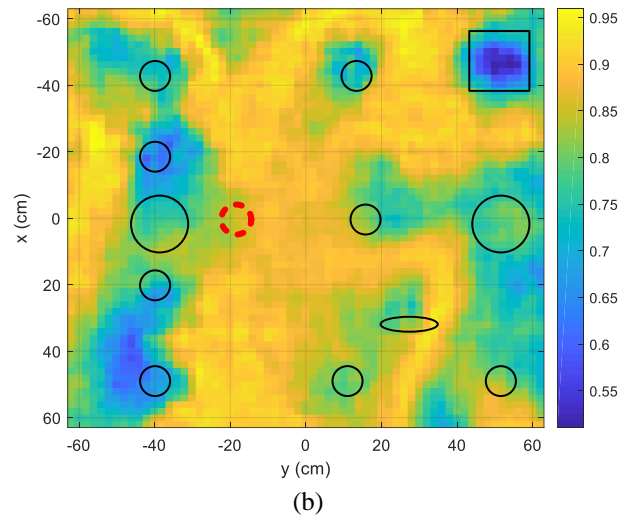

(b)

Fig. 6. Results of the proposed methods. (a) Time gating. (b) Wiener filter. Some buried objects are not detectable in these scenarios. Red dash lines show not detected objects.

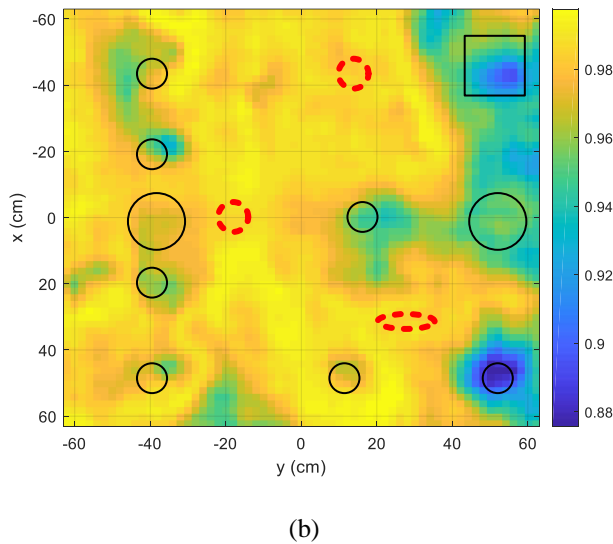

(b)

## V. CONCLUSION

Discontinuities in the environment reflect incident signals back from the boundaries. Discontinuities in the medium can occur due to underground objects being targets of GPRs, but, the ground clutter is a major hurdle while looking for shallow buried objects. In this paper different methods of ground clutter cancellation are compared with each other. Conventional methods have unsuitable results in the clutter cancellation of the shallow buried object problem. Time gating, Wiener filter and the combined wiener filter/time gating methods show better results compared to conventional methods. Finding all buried objects in BOI problem is important task which was improved in comparison to previous investigations in this study. The proposed method of the paper was based on primary detection and finding best time gating window length. Results of time gating method have high rate of precision when suitable window length is determined. Furthermore, results of wiener filter were used for primary detection of targets. Time gating method was then applied with the window length determined on the basis of the previous step results. This combination leads to the best performance among all methods assessed.

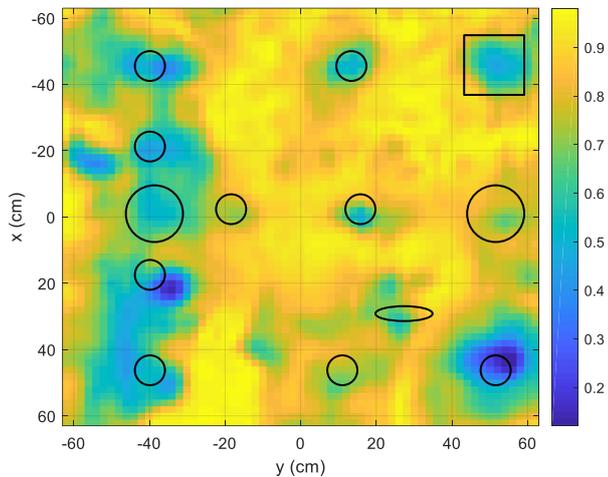

Fig. 7. Results of the novel algorithm based on combined time gating. wiener filter method. All buried objects are detected, although some artifacts are still present